# A Critique of Snapshot Isolation

Daniel Gómez Ferro    Maysam Yabandeh [*]

Yahoo! Research
*Barcelona, Spain*
{danielgf,maysam}@yahoo-inc.com

## Abstract

The support for transactions is an essential part of a database management system (DBMS). Without this support, the developers are burdened with ensuring atomic execution of a transaction despite failures as well as concurrent accesses to the database by other transactions. Ideally, a transactional system provides serializability, which means that the outcome of concurrent transactions is equivalent to a serial execution of them. Based on experiences on lock-based implementations, nevertheless, serializability is known as an expensive feature that comes with high overhead and low concurrency. Commercial systems, hence, compromise serializability by implementing weaker guarantees such as *snapshot isolation*. The developers, therefore, are still burdened with the anomalies that could arise due to the lack of serializability.

There have been recent attempts to enrich large-scale data stores, such as HBase and BigTable, with transactional support. Not surprisingly, inspired by traditional database management systems, serializability is usually compromised for the benefit of efficiency. For example, Google Percolator, implements lock-based snapshot isolation on top of BigTable. We show in this paper that this compromise is not necessary in lock-free implementations of transactional support. We introduce *write-snapshot isolation*, a novel isolation level that has a performance comparable with that of snapshot isolation, and yet provides serializability. The main insight in write-snapshot isolation is to prevent read-write conflicts in contrast to write-write conflicts that are prevented by snapshot isolation.

*Categories and Subject Descriptors*    H.2.4 [*Database Management*]: Systems–concurrency, transaction processing

## 1. Introduction

A transaction is an atomic unit of execution and may contain multiple read and write operations to a given database. A reliable transactional system provides ACID properties: atomicity, consistency, isolation, and durability. *Isolation* defines the system behavior in presence of concurrent transactions. Ideally, the isolation level guarantees serializability, which means that the behavior of the system is equivalent to a system that serially runs the transactions (with no concurrency). Serializability, however, is known to be expensive because of (i) the high implementation overhead, (ii) the lower level of concurrency.

Commercial data storage systems [23, 24], such as Google Percolator [23, 24], hence, often implement a weaker guarantee, *snapshot isolation* [5], since it allows for high concurrency between transactions. In snapshot isolation, the snapshot from which a transaction reads is not affected by the concurrent transactions. To provide read snapshots, the database maintains multiple versions of the data [6] and the transactions observe different versions of the data depending on their start time. Two concurrent transactions still conflict if they write into the same data element, which is known as *write-write* conflict.

One advantage of snapshot isolation is that it checks only for write-write conflicts, which its lock-based implementation [24] is very straightforward: a transaction locks a data item before modifying it and aborts if it is already locked (or waits for the lock to be released). Furthermore, the read-only transactions, which comprise the majority of the transactional traffic, could run without any extra locking overhead since snapshot isolation does not require maintaining locks for reads. The drawback is that serializability, which sometimes requires detecting *read-write* conflicts, is not provided by snapshot isolation. (See Section 7 for the list of approaches for adding serializability to snapshot isolation.)

Adding read-write conflict detection to a lock-based transactional system, however, comes with a non-negligible overhead. This is because the read operations, which are the majority in a typical workload, have to maintain the locks as well. Moreover, in a naive implementation of read-write conflict detection, read-only transactions could be aborted, which would greatly reduce the level of concurrency that the system could provide.

In lock-free implementations of snapshot isolation [20], which is suitable for OLTP traffic, the list of identifiers of modified rows is submitted to a centralized status oracle, where they are checked for write-write conflicts. To check for read-write conflicts instead, the transactions could also submit the identifiers of read rows to the status oracle, to be checked against the modified rows of committed transactions. Therefore, restricting the prevented conflicts to only write-write no longer offers a benefit in terms of implementation overhead. It is time, thus, to revisit the core ideas behind write-write and read-write conflict detections and analyze the guarantees as well as the level of concurrency that they provide.

In this paper, we analyze write-write and read-write conflicts. We show that write-write conflict detection that is provided by snapshot isolation is not necessary for providing serializability. In other words, a system could be serializable and still allow for write-write conflicts. More importantly, we prove that read-write conflict detection is sufficient for providing serializability. Based on this analysis, we see that serializability could be brought into large-scale data stores with an overhead comparable to that of snapshot isolation. We present *write-snapshot isolation*, a new isolation level that prevents read-write conflicts instead of write-write conflicts. Each transaction running under write-snapshot isolation writes into a separate snapshot of the database specified by the transaction commit timestamp. Although a transaction reads from a snapshot specified by its start timestamp, it aborts if the read rows are modified by a concurrent, committed transaction.

We expect the level of concurrency offered by write-snapshot isolation to be comparable with that of snapshot isolation. First, as we show in Section 4, neither write-snapshot isolation nor snapshot isolation aborts read-only transactions, which comprise the majority of transactional traffic [10, 12]. Second, two concurrent (write) transactions that have write-write or read-write conflict could still be serializable, and preventing either one could unnecessarily abort some transactions. Consequently, neither of snapshot isolation and write-snapshot isolation have a clear advantage over the other in terms of the offered level of concurrency, which highly depends on the data access pattern in each application. We therefore leave it to the experimental results to show which isolation level overall offers a higher concurrency.

We have implemented both write-snapshot isolation and snapshot isolation on top of HBase [1], a widely used distributed data store. The experimental results show that the level of concurrency offered by write-snapshot isolation is comparable with that offered by snapshot isolation. Serializability, therefore, could be brought to lock-free transactional systems without, however, hurting the performance.

Main Contributions Here we list the main contributions of this paper:
1. We present an analysis of the core ideas behind snapshot isolation and serializability.
2. We introduce a new isolation level, write-snapshot isolation, that checks for read-write conflicts instead of write-write conflicts.
3. We prove that write-snapshot isolation provides serializability.
4. We present a lock-free implementation of write-snapshot isolation on top of HBase, and show that although write-snapshot isolation provides the precious feature of serializability, it is comparable with snapshot isolation in terms of both the implementation overhead and the offered level of concurrency.

Roadmap The remainder of this paper is organized as follows. Section 2 explains snapshot isolation and overviews both its lock-based and lock-free implementations. Serializability is defined in Section 3, which also analyzes the executions that are allowed under snapshot isolation. Write-snapshot isolation is introduced in Section 4, which is followed by its lock-free implementation presented in Section 5. After evaluating our lock-free implementation of write-snapshot isolation on top of HBase in Section 6, we review the related work in Section 7. This section also gives an overview of the related work in making snapshot isolation serializable [1, 2, 8, 16, 19]. We finish the paper with some concluding remarks in Section 8.

## 2. Snapshot Isolation

Here, we give an overview on both lock-based and lock-free implementations of snapshot isolation. This overview is also presented in our previous work on lock-free implementation of snapshot isolation [20].

Snapshot isolation guarantees that the snapshot from which a transaction reads is not affected by the concurrent transactions. To implement snapshot isolation, the database maintains multiple versions of the data in some *data servers*, and transactions, run by *clients*, observe different versions of the data depending on their start time. Implementations of snapshot isolation have the advantage that writes of a transaction do not block the reads of others. Two concurrent transactions still conflict if they write into the same data item, say a database row.[2] The conflict must be detected by the snapshot isolation implementation, and at least one of the transactions must abort.

---
[1] http://hbase.apache.org

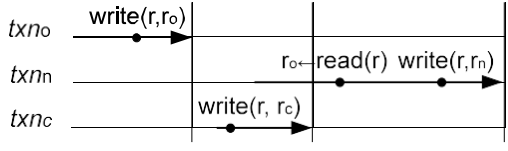

Figure 1. An example run under snapshot isolation guarantee. *write*($r$, $v$) writes value $v$ into data item $r$, and *read*($r$) returns the value in data item $r$. Transaction $txn_n$ observes the commits of transaction $txn_o$ since $txn_o$ commits before $txn_n$ starts. It, however, does not read the writes of transaction $txn_c$, as it is not committed at the time $txn_n$ start timestamp is assigned. Transactions $txn_n$ and $txn_c$ have both spatial and temporal overlap and at least one of them must abort.

To implement snapshot isolation, each transaction receives two timestamps: one before reading and one before committing the modified data. In both lock-based and lock-free approaches, timestamps are assigned by a centralized server, the *timestamp oracle*, and hence provide a commit order between transactions. Transaction $txn_i$ with assigned start timestamp $T_s(txn_i)$ and commit timestamp $T_c(txn_i)$ (denoted [$T_s(txn_i)$, $T_c(txn_i)$]) reads the latest version of data with commit timestamp $\delta < T_s(txn_i)$. In other words, the transaction observes all its own changes as well as the modifications of transactions that have committed before $txn_i$ starts. In the example of Figure 1, transaction $txn_n$ reads the modifications by the committed transaction $txn_o$, but not the ones made by the concurrent transaction $txn_c$.

If $txn_i$ does not have any write-write conflict with another concurrent transaction, it commits its modifications with a commit timestamp. Two transactions $txn_i$ and $txn_j$ conflict if the following holds:

1. Spatial overlap: both write into row $r$;
2. Temporal overlap:
   $T_s(txn_i) < T_c(txn_j)$ and $T_s(txn_j) < T_c(txn_i)$.

In the example of Figure 1, both transactions $txn_n$ and $txn_c$ write into the same row $r$ and therefore conflict (spatial overlap). Since they also have temporal overlap, the snapshot isolation implementation must abort at least one of them.

### 2.1 Lock-based Implementation of Snapshot Isolation

Percolator [24] is a state-of-the-art implementation of this approach on top of a distributed data store. The uncommitted data are written directly into the main database with a version equals to the transaction start timestamp. Percolator [24] adds two extra columns to each column family: *lock* and *write*. The write column maintains the commit times-tamp. The client runs a 2PC algorithm to update this column on all modified data items. The lock columns provide low granularity locks to be used by the 2PC algorithm. In the first phase of 2PC, the client writes the data and acquires the corresponding locks. Depending on the implementation, if a transaction tries to write into a locked data, it could (i) wait on the lock, (ii) abort, or (iii) force the abort of the transaction that is holding the lock. In the second phase of 2PC, the client updates the data with the commit timestamp and removes the locks. Although using locks simplifies the write-write conflict detection, the locks a failed or slow transaction holds prevent the others from making progress during recovery.

---
[2] Here, we use the row-level granularity to detect the write-write conflicts. It is possible to consider finer degrees of granularity, but investigating it further is out of the scope of this work.

### 2.2 Lock-free Implementation of Snapshot Isolation

In the lock-free implementation of snapshot isolation, a single server, *i.e.*, the status oracle, receives the commit requests accompanied by the set of the identifiers of modified rows, $R$. Since status oracle has observed the modified rows by the previous commit requests, it could maintain the commit data and therefore has enough information to check the temporal overlap condition for each modified row. Efficient implementations of this approach could service up to 50K TPS [20] (where each transaction modifies 10 rows on average), which shows that the status oracle is not a bottleneck for scalability of the system. Appendix A explains how the related work [20] addresses the challenges in implementing the status oracle in an efficient and reliable manner. Appendix A presents a brief overview of the techniques presented in our previous work [20] to address the challenges in implementing the status oracle in an efficient and reliable manner.

Algorithm 1 describes the procedure to process a commit request for a transaction $txn_i$. In the algorithm, $R$ is the list of all the modified rows, $T_c$ is the state of status oracle containing the commit timestamp of transactions, and *lastCommit* is the state of status oracle containing the last commit timestamp of the modified rows.

---

**Algorithm 1** Commit request ($T_s(txn_i)$, $R$) : {commit, abort}

1: **for each** row $r \in R$ **do**
2:     **if** *lastCommit*($r$) > $T_s(txn_i)$ **then**
3:         **return** abort;
4:     **end if**
5: **end for**
                                                                  ◁ Commit $txn_i$
6:  $T_c(txn_i) \rightarrow$ TimestampOracle.next();
7: **for each** row $r \in R$ **do**
8:     *lastCommit*($r$) $\rightarrow T_c(txn_i)$;
9: **end for**
10: **return** commit;

---

To check for write-write conflicts, Algorithm 1 checks temporal overlap for all the already committed transactions. In other words, in the case of a write-write conflict, the algorithm commits the transaction for which the commit re-

157

quest is received sooner. Temporal overlap condition must be checked on every row $r$ modified by transaction $txn_i$ against all the committed transactions that have modified the row. Line 2 performs this check, but only for the latest committed transaction $txn_l$ that has modified row $r$. We can show by induction that this check guarantees that temporal overlap condition is respected by all the committed transactions that have modified row $r$. Also, notice that Line 2 verifies only the first part of temporal overlap property. This is sufficient in the status oracle because the commit timestamps are obtained by status oracle in contrast to the general case in which commit timestamp could be obtained by clients [24]. Line 6 maintains the mapping between the transaction start and commit timestamps. This data could be used later to process queries about the transaction statuses [20].

To obtain the read snapshot of a transaction, the transaction compares its start timestamp with the commit timestamp of the written values. The reading transaction $txn_r$ skips a particular version, if the transaction $txn_w$ that has written it is (i) not committed yet, (ii) aborted, or (iii) committed with a commit timestamp larger than the start timestamp of $txn_r$. The commit timestamps could in general be obtained form the status oracle server. Alternatively, to avoid additional calls into the status oracle server, depending on the implementation, they could be written back into the database [20] or be replicated on the clients [17]. The experiments in this paper is performed on an implementation based on the latter approach.

## 3. Serializability

A *history* represents the interleaved execution of transactions as a linear ordering of their operations [5]. To show the histories, we use the notation presented in [5]: "w1[x]" and "r1[x]" denotes a write and a read by transaction $txn_1$ on data item x, respectively. Commits and aborts of $txn_1$ are shown by "c1" and "a1", respectively. A history is *serial* if its transactions are not concurrent. Two histories are equivalent if they include the same transactions and produce the same output.

### 3.1 Is Write-write Conflict Avoidance *Sufficient* for Serializability?

Snapshot isolation is not serializable [5], which means that it allows histories that do not have serial equivalence. For example, if transaction $txn_1$ reads x and writes y and transaction $txn_2$ reads y and writes x, then the following history is possible under snapshot isolation:

H 1. *r1[x] r2[y] w1[y] w2[x] c1 c2*

The snapshot isolation implementation does not prevent History 1 since the transactions write into different data items, i.e., do not have spatial overlap. This could lead to a well-known anomaly called *write skew* [5]. The practical problem that write skew could arise is that the write set of the interleaving transactions could be related by a *constraint* in the database. Even if each transaction validates the constraint before its commit, two concurrent transactions could still violate the constraint. For example, assume the constraint of $x + y > 0$ and initial values of $x = y = 1$. Further assume that transaction $txn_1$ reads x and y, and decreases x by one if the constraint condition is still valid. Transaction $txn_2$ does the same but decreases from y. Snapshot isolation allows the following history:

H 2. *r1[x] r1[y] r2[x] r2[y] w1[x] w2[y] c1 c2*

History 2 is not serializable and transforms the database into the state of $x = y = 0$, which violates the constraint.

### 3.2 Is Write-write Conflict Avoidance *Necessary* for Serializability?

Although snapshot isolation is not serializable, it prevents many anomalies in data, including the ones listed in the ANSI SQL Standard [3]: (i) *dirty read*: reading an uncommitted value, (ii) *fuzzy read*: having an already read value deleted by a concurrent transaction, and (iii) *phantom*: the set of items that satisfy a search condition vary due to modifications made by concurrent transactions. Snapshot isolation does not have this problem since it reads from a snapshot of the database that is not affected by concurrent transactions. Note that this is independent of the particular conflict detection mechanism, which is write-write conflict detection here, and these anomalies do not manifest even if we do not prevent any kind of conflicts. Besides the ANSI-listed anomalies, it prevents the *Lost Update* anomaly [5], in which the updates of a committed transaction are lost after the commit of a concurrent transaction. For example, in the following unserializable history the updated value x by transaction $txn_1$ is lost after commit of transaction $txn_2$.

H 3. *r1[x] r2[x] w2[x] w1[x] c1 c2*

Snapshot isolation prevents History 3 because both transactions write to x and therefore have write-write conflict. Note that in History 3 if transaction $txn_2$ does not read x (i.e., blind write to x), such as in History 4, the lost update anomaly does not manifest.

H 4. *r1[x] w2[x] w1[x] c1 c2*

This is because the history is equivalent to the following serial history:

H 5. *r1[x] w1[x] c1 w2[x] c2*

After the execution of both histories, x is updated by the write of $txn_2$, i.e., w2[x]. The modifications made by transaction $txn_1$ are updated by $txn_2$, but they are certainly *not lost*. They are visible by any transaction $txn_k$ with a start timestamp between the two commits: $T_c(txn_1) < T_s(txn_k) < T_c(txn_2)$. The lost update anomaly was vaguely explained in [5] (by mentioning the read of transaction $txn_2$ in parenthesis), which could give the wrong impression that avoiding write-write conflicts is always necessary. Quite the contrary, avoiding write-write conflicts could unnecessarily prevents some serializable histories such as History 4. In other words, write-write conflict avoidance of snapshot isolation, besides allowing some histories that are not serializable, unnecessarily lowers the concurrency of transactions by preventing some valid, serializable histories.

## 4. Read-Write vs. Write-Write

Multi-version databases [6] (MVCC) maintain multiple versions for the data and add the new data as a new version instead of rewriting the old data. This enables the transactions to read from an arbitrary snapshot of the database (usually specified by the transaction start timestamp) and write to an arbitrary snapshot (specified by the transaction commit timestamp). To implement optimistic concurrency control [21] on top of a multi-version database, further checks must be performed at the commit time.

Snapshot isolation adds write-write conflict detection to MVCC. In fact, snapshot isolation could be termed *read-snapshot isolation* since the read phase of a transaction is never interrupted by concurrent transactions, i.e., the *read* snapshot is isolated. Instead of write-write conflict detection, *write-snapshot isolation* adds read-write conflict detection to MVCC, which means that a transaction does not commit if its read set is modified by a concurrent transaction. However, in contrast with read-snapshot isolation, the write phase of a transaction running under write-snapshot isolation is never interrupted by concurrent transactions, i.e., the *write* snapshot is isolated. We showed in Section 3 that (read-) snapshot isolation prevents some histories that are serializable (e.g., History 4) and allows some that are not (e.g., History 2). In this section, we formally define write-snapshot isolation and the guarantees it provides.

### 4.1 Write-Snapshot Isolation

As we explained, multi-version databases enable the transactions to operate on separate snapshots, where snapshots are composed of different versions of data. The precise definition of the snapshot from which a transaction $txn_i$ reads depends on the implementation of the isolation level. Similarly to snapshot isolation, write-snapshot isolation assigns unique start and commit timestamps to transactions and ensures that $txn_i$ reads the latest version of data with commit timestamp $\delta < T_s(txn_i)$. In other words, the transaction observes all its own changes as well as the modifications of transactions that have committed before $txn_i$ starts. The difference between write-snapshot isolation and snapshot isolation is, however, in the way the conflict between two transactions is defined.

Formally speaking, two transactions $txn_i$ and $txn_j$ conflict under write-snapshot isolation if the following holds:

1. RW-spatial overlap: $txn_j$ writes into row $r$ and $txn_i$ reads from row $r$;

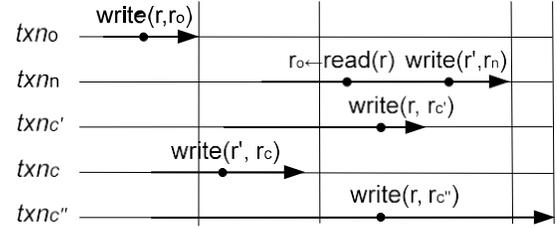

Figure 2. An example run under write-snapshot isolation guarantee. $write(r, v)$ writes value $v$ into data item $r$, and $read(r)$ returns the value in data item $r$. Transaction $txn_n$ observes the commits of transaction $txn_o$ since $txn_o$ commits before $txn_n$ starts. It, however, does not read the writes of transaction $txn_{c'}$ as it is not committed at the time $txn_n$ start timestamp is assigned. Transactions $txn_n$ and $txn_{c'}$ have both rw-spatial and rw-temporal overlap and at least one of them must abort. Although transactions $txn_n$ and $txn_c$ have rw-temporal overlap, they do not have read-write conflict since $txn_c$ does not write into row $r$ that is in the read set of $txn_n$. Similarly, $txn_n$ and $txn_{c''}$ do not have conflict because they do not have rw-temporal overlap.

2. RW-temporal overlap:
   $T_s(txn_i) < T_c(txn_j) < T_c(txn_i)$.

In other words, transactions that commit during the lifetime of transaction $txn_i$ should not modify its read data.

Note: The definition of rw-temporal overlap is different from the temporal overlap explained in Section 2. For example, although transactions $txn_n$ and $txn_{c''}$ in Figure 2 have temporal overlap under snapshot isolation, they do not have rw-temporal overall under write-snapshot isolation. This is because $txn_{c''}$ that modifies the read data of $txn_n$ does not commit during the lifetime of transaction $txn_n$. Since transaction $txn_n$ does not modify the read data of transaction $txn_{c''}$ (which is empty here), its commit time being during the lifetime of transaction $txn_{c''}$ does not cause an rw-temporal overlap.

In the example of Figure 2, transaction $txn_{c'}$ writes into the same row from which $txn_n$ has read (rw-spatial overlap). Since they also have rw-temporal overlap, the write-snapshot isolation implementation must abort at least one of them. However, although transactions $txn_n$ and $txn_c$ write into the same row $r'$, they do not have rw-spatial overlap under write-snapshot isolation since transaction $txn_c$ does not modify the read data of transaction $txn_n$, i.e., row $r$. Two concurrent transactions $txn_n$ and $txn_{c''}$ do not have rw-temporal overlap (because $T_c(txn_{c''}) > T_c(txn_n)$) and therefore are allowed under write-snapshot isolation.

*Read-only transactions* We use the term "*write transaction*" to refer to a transaction in which the write set is not empty. A transactions is *read-only* if its write set is empty. These transactions are important since they constitute a large part of transactional traffic. For example, in TPC-E [12] benchmark around 77% of transactions are read-only [10], and efficient support for them have a huge impact on the overall performance. Moreover, Google Megastore [4], which services 23 billion transactions daily on top of a key-value store [3], reports more than 86% share for read-only transactions. It is, therefore, very important to ensure that (i) the overhead of running read-only transactions under write-snapshot isolation is close to a minimum, and (ii) the read-only transactions never abort under write-snapshot isolation. We will show in Section 5.1 that the sole overhead of write-snapshot isolation for read-only transactions is obtaining the start timestamp, the same overhead as in snapshot isolation. Here, we show how to avoid abort of read-only transactions in write-snapshot isolation.

Plainly, since a read-only transaction does not perform any writes, it does not affect the values read by other transactions, and therefore does not affect the concurrent transactions as well. Because the reads in both snapshot isolation and write-snapshot isolation are performed on a fixed snapshot of the database that is determined by the transaction start timestamp, the return value of a read operation is always the same, independent of the real time that the read is executed. Hence, a read-only transaction is not affected by concurrent transactions and intuitively does not have to be aborted. As depicted in Figure 3, a read transaction $[T_s(txn_r), T_c(txn_r)]$ is equivalent to transaction $[T_s(txn_r), T_c]$, where $T_s(txn_r) \le T_c$. We, therefore, optimize the definition of read-write conflict in write-snapshot isolation by adding the following condition:

3. Not read-only: none of transactions $txn_i$ and $txn_j$ is read-only.

In other words, the read-only transactions are not checked for conflicts and hence never abort.

### 4.2 Is Read-write Conflict Avoidance *Sufficient* for Serializability?

Here we prove that write-snapshot isolation is serializable. To this aim, we need to show that each history, $h$, run under write-snapshot isolation is *equivalent* to a *serial* history *serial*($h$) [5]. To keep two histories *equivalent*, we keep the same order for (i) operations inside a transaction and (ii) transaction commits. In this way, if a transaction in the new history reads from the same snapshot as in the original history, it commits the same values as well. One way to achieve that is to keep the same order for transaction starts as well. In this way, a transaction observes the same history of commits and, therefore, reads from the same snapshot as in the original history. However, to have the new history *serial*, we must avoid overlapping between transactions. We do that by shifting operations of write (resp. read-only) transactions to the commit (resp. start) point. Intuitively, because write-snapshot isolation prevents read-write conflicts between write transactions, the shifting does not affect the observed commits by transactions.

---
[3] Megastore achieves this performance by sacrificing serializability. It partitions the data store, and provides limited consistency guarantees across partitions. See Section 7 for further details.

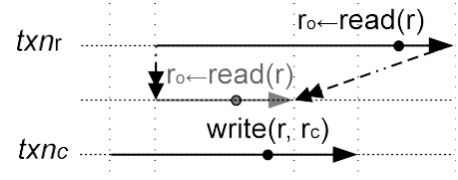

Figure 3. Each read-only transaction run under write-snapshot isolation is equivalent to a shorter transaction with the same start timestamp. This is because the read operations are serviced from a snapshot of the database, and the real time of performing the read does not affect the return value.

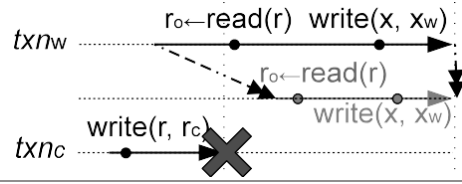

Figure 4. Each write transaction run under write-snapshot isolation is equivalent to a shorter transaction with the same commit timestamp. This is because the read set of a write transaction is not modified by any transaction with rw-temporal overlap.

Putting it together, we construct history *serial*($h$) by:

1. Using the same commit order of history $h$ for write transactions;
2. Maintaining the order of operations inside each transaction;
3. Moving all the operations of a read-only transaction to right after its start.
4. Moving all the operations of a write transaction to right before its commit.

The aborted transactions could be excluded since, similarly to snapshot isolation, their modifications are not read by other transactions.

Lemma 1. *History serial(h) is serial.*

*Proof.* Since all operations of each transaction are either right before its commit or right after its start, and the assigned timestamps are unique, there are no concurrent transactions in *serial*($h$), and it is, therefore, serial. □

Lemma 2. *History serial(h) is equivalent to history h.*

*Proof.* As we explained before, the values read by a read-only transaction change by neither the real time of the read operations nor the commit time. Since the commit timestamp of write transactions is preserved, by using the same start timestamp a read-only transaction observes the same commits in its read snapshot, and hence read the same values. The output of a history is determined by the commit of write transactions. Since the commit order of write transactions is preserved in history *serial(h)*, the output is the same as that of history *h*, as long as the read values by each write transaction is the same. As depicted in Figure 4, this is the case since read-write conflicts do not manifest in a write-snapshot isolation history. In other words, a write transaction $[T_s(txn_i), T_c(txn_i)]$ is equivalent to transaction $[T_s, T_c(txn_i)]$, where $T_s(txn_i) \leq T_s \leq T_c(txn_i)$. This is because the read set of write transaction $txn_i$ is not modified by any transaction that is committed during the lifetime of $txn_i$. □

Theory 1. *write-snapshot isolation is serializable.*

*Proof.* Based on Lemmas 1 and 2, for each history *h* run under write-snapshot isolation, we can construct a serial history *serial(h)*, which is equivalent to history *h*. □

We showed that for each write-snapshot isolation history, we can obtain a serial-equivalent history in which the transactions are ordered according to their commit timestamp. Since write-snapshot isolation is serializable, it does not allow the anomalies specified by the ANSI SQL Standard [3] (which are avoided by snapshot isolation) as well as the anomalies that could manifest under snapshot isolation. For example, in History 1 $txn_1$ (i) commits during the lifetime of $txn_2$, and (ii) writes into y from which $txn_2$ has read, and one of them, therefore, must abort. Also, in History 2, which is an example of write skew, $txn_1$ that commits sooner, writes into x from which $txn_2$ reads and they, hence, conflict. Moreover, the lost update anomaly that is prevented by snapshot isolation is also prevented by write-snapshot isolation. For example, in History 3 $txn_1$ (i) writes into x from which $txn_2$ has read, and (ii) commits during the lifetime of $txn_2$, and therefore has read-write conflict with $txn_2$.

### 4.3 Is Read-write Conflict Avoidance *Necessary* for Serializability?

One advantage of write-snapshot isolation over snapshot isolation is that the concurrent transactions that are unnecessarily aborted due to a write-write conflict in snapshot isolation are allowed in write-snapshot isolation. For example, write-snapshot isolation allows serializable History 4 because (i) $T_c(txn_1) < T_c(txn_2)$, and (ii) $txn_1$ does not write into the read set of $txn_2$ (which is empty). Nevertheless, some other serializable histories are unnecessarily prevented by write-snapshot isolation as well. For example, consider the following history:

H 6. *r1[x] r2[z] w2[x] w1[y] c2 c1*

In this history, after commit c2 the new value of x is updated based on the value of z, and after commit c1 the value of y is updated based on the old value of x that was read before commit c2. Write-snapshot isolation prevents History 6 because transaction $txn_2$ that commits during lifetime of transaction $txn_1$ writes into x from which $txn_1$ has read. However, the history is serializable as shown in the following history:

H 7. *r1[x] w1[y] c1 r2[z] w2[x] c2*

After running serial History 7, the value of y is updated based on the old value of x, and the new value of x is updated based on the value of z, which is the same output as History 6.

Write-snapshot isolation has the advantage of offering serializability, the precious feature that snapshot isolation is missing. Both snapshot isolation and write-snapshot isolation unnecessarily abort some serializable transactions. The rate of unnecessary aborts highly depends on the particular workload under which the system runs. We, therefore, leave it to the experimental results to show that overall which isolation level offers a higher concurrency.

## 5. Lock-free Implementation of Write-snapshot Isolation

Here, we present a lock-free implementation of write-snapshot isolation and show that in the lock-free scheme, the overhead of snapshot isolation and write-snapshot isolation are comparable.

Similar to the lock-free implementation of snapshot isolation presented in Section 2, we use the status oracle server to commit the transaction. The status oracle maintains the list of identifiers of modified rows by committed transactions. Each commit request comprises two sets: the set of identifiers of modified rows, $R_w$, and the set of identifiers of read rows, $R_r$. The read set is checked against the modified rows of concurrent committed transactions. If there is no read-write conflict, the status oracle commits the transaction and uses the write set to update the list of modified rows in the status oracle. Note that the set of identifiers of the read rows that is submitted to the status oracle is computed based on the rows that are actually read by the transaction, whether these rows were originally specified by their primary keys or by a search condition.

Algorithm 2 describes the procedure to process a commit request for a transaction $txn_i$. In the algorithm, $R_w$ is the list of all the modified rows, $R_r$ is the list of all the read rows, $T_c$ is the state of status oracle containing the commit timestamp of transactions, and *lastCommit* is the state of status oracle containing the last commit timestamp of the modified rows. Similar to Algorithm 1, Line 2 performs the rw-spatial check only for the latest committed transaction $txn_l$ that has modified row $r \in R_r$. Notice that here we check for the read rows $R_r$ in contrast with the write rows in Algorithm 1. Af-

**Algorithm 2** Commit request ($T_s(txn_i)$, $R_w$, $R_r$) : {commit, abort}

1: for each row $r \in R_r$ do
2:     if $lastCommit(r) > T_s(txn_i)$ then
3:         return abort;
4:     end if
5: end for
                                          ◁ Commit $txn_i$
6: $T_c(txn_i) \rightarrow$ TimestampOracle.next();
7: for each row $r \in R_w$ do
8:     $lastCommit(r) \rightarrow T_c(txn_i)$;
9: end for
10: return commit;

ter committing the transaction, Line 8 updates the *lastCommit* state by the write set $R_w$. As we can see, the changes into the implementation of snapshot isolation presented in Section 2 are a few and the overhead of lock-free implementations of snapshot isolation and write-snapshot isolation are comparable. The commit request is a little bigger in write-snapshot isolation since it also includes set $R_r$. However, since status oracle is a CPU-bound service [17, 20], the network interface bandwidth of the status oracle server is greatly under-utilized and slightly larger packet sizes do not affect its performance.

### 5.1 Read-only Transactions

We showed in Section 4 that neither of write-snapshot isolation and snapshot isolation aborts the read-only transactions. Here we show that the centralized implementations of write-snapshot isolation and snapshot isolation impose the same overhead for read-only transactions.

Since a read-only transaction always commits in both write-snapshot isolation and snapshot isolation, the client does not have to submit any value with the commit request and the status oracle server does not pay the cost of processing the commit request. This is naturally followed in snapshot isolation since a read-only transaction in snapshot isolation submits an empty list of written rows with its commit request [4]. Therefore, according to Algorithm 1 the transaction always commits since there is no write-write conflict with other transactions. To implement this feature in write-snapshot isolation, the client submits an empty read set to the status oracle if its write set is empty (i.e., is read-only). According to Algorithm 2, therefore, since both read and write sets are empty, the status oracle commits without performing any computation for the transaction.

### 5.2 Analytical Traffic

The lock-free implementation using a centralized status oracle [20] is designed for online transaction processing (OLTP), which is typically composed of small, short trans-actions. Analytical traffic, which could include transactions with a very large read set, is out of the scope of this paper. For example, a transaction could scan the entire database and compute some statistics over a field. To illustrate the possible future work, here we mention the two main challenges in extending this implementation for efficient support of occasional analytical traffic. First, the read set could become very large and submitting that to the status oracle could be expensive. Second, the larger the read set, the higher is the probability of a read-write conflict and thus the higher is the abort rate. To address the former, analytical transactions could submit to the status oracle a compact, over-approximated representation of the read set, e.g., table name and row ranges. The latter challenge, which is more fundamental, could be addressed by treating the analytical transactions differently. For example, if a mechanism could ensure that the computed statistics by the analytical traffic are not used by OLTP transactions, which is normally the case, their commit will not affect the OLTP traffic and could be entirely skipped.

---
[4] In general, it is possible to completely drop the commit request of a read-only transaction, but it will incur some recovery cost on the status oracle since it suspects the failure of the client [20].

## 6. Evaluation

Here we compare the concurrency level offered by a centralized, lock-free implementation of write-snapshot isolation with that of snapshot isolation presented in [20]. We have implemented two prototypes that integrate write-snapshot isolation (WSI) and snapshot isolation (SI) with HBase, a clone of Bigtable [9] that is widely used in production applications. HBase provides a scalable key-value store, which supports multiple versions of data. It splits groups of consecutive rows of a table into multiple regions, and each region is maintained by a single data server (RegionServer in HBase terminology). A transaction client has to read/write cell data from/to multiple regions in different data servers when executing a transaction. To read and write versions of cells, clients submit get/put requests to data servers. The versions of cells in a table row are determined by timestamps [20].

We used 34 machines with 2.13 GHz Dual-Core Intel(R) Xeon(R) processor, 2 MB cache, and 4 GB memory: 1 for the ZooKeeper coordination service [18], 2 for Book-Keeper [5], 1 for status oracle, 25 for data servers, and 5 for hosting clients. BookKeeper is a system to perform write-ahead logging efficiently and reliably: every change into the memory of the status oracle that is related to a transaction commit/abort is persisted in multiple remote storages via BookKeeper. ZooKeeper is a coordination service that is used by both HBase and BookKeeper. HBase is initially loaded with a table of size 100 GB comprising 100M rows. Since the allocated memory to each HBase process is 3 GB, this table size ensures that the data does not fit into the memory of data servers, representing a system operating on very large-scale data. A random read, therefore, causes an IO

---
[5] http://zookeeper.apache.org/bookkeeper

operation from either a local or remote hard disk. The evaluations aim to answer the following questions:

1. What is the overhead of checking for read-write conflicts in write-snapshot isolation compared to checking for write-write conflicts in snapshot isolation?

2. What is the level of concurrency offered by write-snapshot isolation compared to that of snapshot isolation?

### 6.1 Benchmark

Ideally, the centralized implementation of write-snapshot isolation and snapshot isolation should be benchmarked with a standard application, generating a typical workload representing the behavior of practical systems. However, transactional support is a new feature to large data stores [20, 24] and the applications that are adapted to use transactions are being developed. Well-established benchmarks such as TPC-E [12], also, have the problem of being designed for SQL databases rather than key-value stores, for which centralized, lock-free implementations of snapshot isolation are developed [20]. We, therefore, use the Yahoo! Cloud Serving Benchmark, YCSB [11], which is a framework for benchmarking large key-value stores. The vanilla implementation operates on single rows and thus does not support transactions. We modified YCSB to add support for transactions, which touch multiple rows. We defined two types of transactions:

1. *Read-only*: where all operations are only read.
2. *Complex*: consists of 50% read and 50% write operations.

Each transaction operates on $n$ rows, where $n$ is a uniform random number between 0 and 20. Based on these types of transactions, we define a *complex* workload, consisting of only complex operations, and a *mixed* workload consisting of 50% read-only and 50% complex transactions.

### 6.2 Microbenchmarks

Here we run the system with one client and break down the latency of different operations involved in a transaction: (i) start timestamp request, (ii) read, (iii) write, and (iv) commit request. The commit latency is measured from the moment that the commit request is sent to status oracle until when its response is received. The average commit latency is 4.1 ms, which is mainly contributed by persistent storage of the commit data into the WAL via BookKeeper. The average latency of start timestamp request is 0.17 ms. Although the assigned start timestamps must also be persisted, the timestamp oracle could reserve thousands of timestamps per each write into the write-ahead log, and therefore on average servicing timestamps does not inflict a persistence cost.

Each random read and write into HBase takes 38.8 ms and 1.13 ms on average, respectively. The writes are in general less expensive since they usually include only writing into memory and appending into a write-ahead log. Random reads, on the other hand, might inflict the cost of loading an entire block from HDFS (the distributed file system used by HBase), and therefore have higher delays.

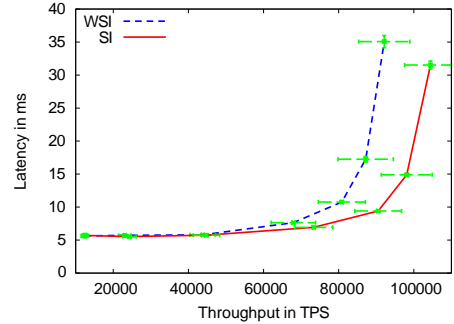

Figure 5. Overhead on the status oracle.

### 6.3 Overhead on the status oracle

The complexity of the commit algorithm in the status oracle is very similar in both snapshot isolation and write-snapshot isolation and we do not expect a big difference in the performance that the status oracle delivers. To measure the relative overhead of snapshot isolation and write-snapshot isolation on the status oracle, here we evaluate both snapshot isolation and write-snapshot isolation on a recent implementation of the status oracle [17]. To stress the status oracle, we need to generate a large volume of traffic, which requires thousands of HBase servers. We therefore evaluate the status oracle in isolation from HBase, and leave measuring the overhead on HBase for the next experiment. This allows the clients to emulate thousands of transactions and stress the status oracle under a high load. Each client allows for 100 outstanding transactions with the execution time of zero, which means that the clients keep the pipe on the status oracle full. We exponentially increase the number of clients from 1 to $2^6$ and plot the average latency vs. the average throughput in Figure 5. The read-only transactions do not cause to the status oracle the cost of checking for conflicts as well as the cost of persisting data into the WAL. To evaluate the write-snapshot isolation performance under a high load, we, therefore, use a *complex* workload where rows are randomly selected out of 20M rows. [6]

As Figure 5 depicts, by increasing the load on the status oracle, the throughput with write-snapshot isolation increases up to 80K TPS with average latency of 10.7 ms. After this point, with increasing the load the latency increases (mostly due to the buffering delay at the status oracle) with only marginal improvement in throughput (92K TPS). Although the difference between the performance of status oracle with snapshot isolation and write-snapshot isolation is negligible when status oracle is not overloaded, status oracle eventually saturates sooner with write-snapshot isolation than with snapshot isolation (104K TPS). The reason is that for the sake of simplicity, the current implementation of status oracle executes the conflict detection algorithm in a critical section. The running time of the critical section is slightly higher with write-snapshot isolation since it requires loading as twice memory items as with snapshot isolation. While write-snapshot isolation loads some memory items to check against the read set and after commit loads some others to update with the write set, snapshot isolation updates the same memory items that are already loaded into the processor cache for write-write conflict detection. Although, this does not cause a tangible increase in processing the individual commit requests, under a heavy load it makes the system be saturated sooner. For future work, we are considering using smaller critical sections to alleviate this issue both for snapshot isolation and write-snapshot isolation. Note that it is not advisable to use a system in its saturation point, and therefore the difference between snapshot isolation and write-snapshot isolation remains negligible under a normal load.

---

[6] Note that the complex workload is different from the write-only workload, for which we reported the throughput of 50K TPS in our previous work [20]. Moreover, the reported performance is for one status oracle implemented on a simple dual-core machine. To get a higher throughput, one could partition the database and use a status oracle for each partition.

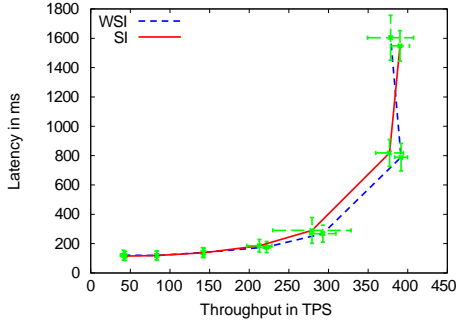

Figure 6. Performance with normal distribution.

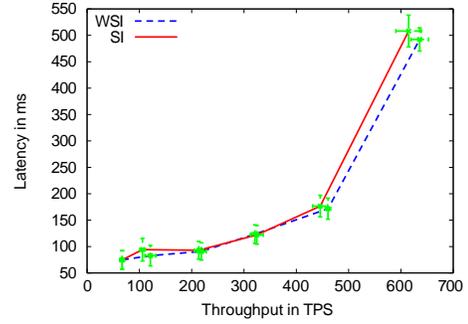

Figure 7. Performance with zipfian distribution.

### 6.4 Overhead on HBase

To compare the overhead of supporting snapshot isolation and write-snapshot isolation, we increase the number of clients from 5 to 10, 20, 40, 80, 160, 320, 640, and plot the average latency vs. the average throughput in Figure 6. The throughput indicates the concurrency level and the difference between the latencies of the two isolation levels compares their relative overhead. The client runs one transaction at a time, where each transaction updates $n$ rows, randomly selected with a uniform distribution on 20M rows. The uniform distribution of rows evenly distributes the load on all the data servers. Therefore, the probability of accessing the same row by two transactions is low and the abort rate will be close to zero. Since almost no transaction is aborted by either write-snapshot isolation or snapshot isolation, the results of this experiments emphasis the overhead of checking the conflicts in write-snapshot isolation and snapshot isolation, excluding the level of concurrency that they could offer.

As we expected from the analysis in Section 5, the overhead of supporting two isolation levels is almost the same and both write-snapshot isolation and snapshot isolation have almost the same performance. After 320 clients, the HBase servers saturate with 391 TPS in write-snapshot isolation. At this point adding more clients does not improve the throughput and increases only the latency due to queuing delays.

### 6.5 Concurrency

To assess the offered concurrency, we repeat the same experiments but with *zipfian* and *zipfianLatest* distributions for selecting the rows. Zipfian distribution models the use cases in which some items are extremely popular [11]. The popular items in zipfianLatest distribution are among the recently inserted data. The high frequent access to popular items increases the probability of conflict between two transactions, and therefore challenges the concurrency level offered by the isolation levels.

Figure 7 depicts the performance under zipfian distribution. Because with this distribution most of the traffic operates on a small proportion of data, random reads are most likely to be serviced from the data already loaded into data servers. Therefore, we see a better throughput and lower latency compared to experiments with a uniform distribution. After 160 clients, however, the cost of processing messages saturates the data servers and adding more clients largely increases the latency, with only marginal improvement on throughput. At this point, the throughput of write-snapshot isolation is 461 TPS and the latency is 172 ms. Overall, the performance of write-snapshot isolation is comparable to that of snapshot isolation.

Figure 8 plots the average abort rate vs. the average throughput for zipfian distribution. The abort rate linearly increases with the increase of throughput, up to 20% in write-snapshot isolation. Although the abort rate in write-snapshot isolation is slightly higher than in snapshot isolation, the difference is negligible. This shows that both snapshot isolation and write-snapshot isolation offer the same level of concurrency for the mixed workload with zipfian distribution.

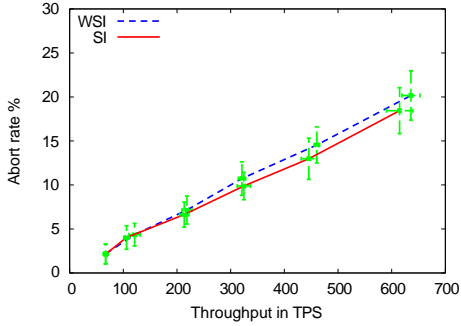

Figure 8. Abort rate with zipfian distribution.

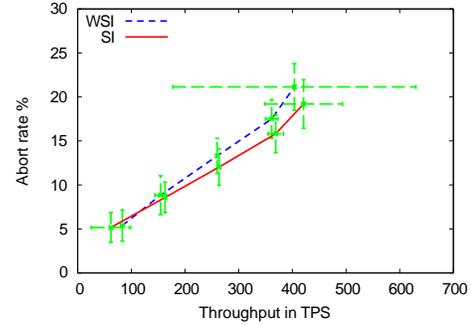

Figure 10. Abort rate with zipfianLatest distribution.

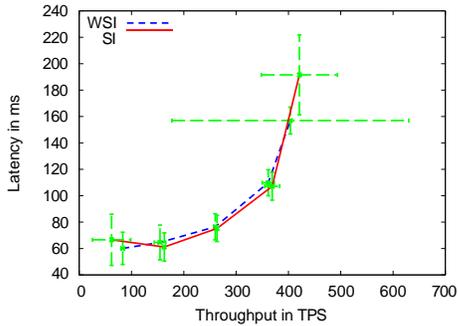

Figure 9. Performance with zipfianLatest distribution.

Figure 9 depicts the performance under zipfianLatest distribution. The performance in this distribution is in general less than in zipfian distribution. Both write-snapshot isolation and snapshot isolation saturate at 40 clients, where the throughput of write-snapshot isolation is 361 TPS and the latency is 110 ms. Nevertheless, the two systems offer a very similar performance. Figure 10 illustrates the abort rate with this distribution. The abort rate with zipfianLatest increases more quickly compared to zipfian. Although the abort rates are similar in write-snapshot isolation and snapshot isolation, it is slightly larger in write-snapshot isolation: with throughput of 361 TPS the abort rate under write-snapshot isolation is 21%, which is 2% larger than that under snapshot isolation. This is because in zipfianLatest the read set is selected mostly from the recent written data, which increases the chance of a read-write conflict in write-snapshot isolation. This slight overhead is the cost that we pay to benefit from the serializability feature offered by write-snapshot isolation.

## 7. Related Work

Here we review the related work in literature. First, we contrast our work with related research in isolation levels. Then, we review the recent implementations of snapshot isolation in large-scale data stores. Finally, we list the large-scale data stores that provide some level of serializability.

### 7.1 Isolation Level

Some previous works [7, 15, 21], check for both read-write and write-write conflicts to provide serializability. Optimistic Concurrent Control (OCC) suggests optimistically running the transactions and postponing the check for conflicts to the commit time [21]. To provide serializability, Kung and Robinson [21] present two lock-based algorithms: one inefficient algorithm with serial validation that hold a write-lock on the whole database (which essentially avoids write-write conflicts), and one efficient algorithm with parallel validation that checks for both read-write and write-write conflicts. Write-snapshot isolation is an implementation of optimistic concurrency control, but further assumes an underlying multi-version database and provides serializability by performing only read-write checks. Moreover, the definition of a read-write conflict (presented in Section 4) is different from that presented in [21].

In the field of transactional memory, TL2 [15] is a lock-based algorithm that checks for both read-write and write-write conflicts. Holding locks on write elements performs the latter: a transaction aborts if it cannot acquire all the locks on the write set. A more recent work that provides one-copy serializability for a replicated database system [7] uses a centralized certifier, which is similar to the status oracle presented in this paper, to check for both read-write and write-write conflicts. One main contribution of this paper is to show that read-write conflict detection is sufficient for serializability, which could be efficiently implemented in lock-free approaches.

There have been several attempts to serialize snapshot isolation. Some are based on a static analysis on the application source code to detect potential conflicting transactions under snapshot isolation [16, 19] and to suggest changes in the application to avoid those conflicts [16]. The modifications proposed by [16] essentially translate the potential conflicts into some write-write conflicts, which are detectable by snapshot isolation. These approaches cannot apply to dynamically generated transactions and require the de-

165

veloper's knowledge on the semantics of snapshot isolation. Our approach is serializable and does not demand any expert knowledge of the developers. Moreover, our approach detects the conflict at the database level and hence does not depend on the application source code. One advantage of write-snapshot isolation over the above, complicated approaches is its simplicity: by slight modifications into the snapshot isolation implementation, write-snapshot isolation adds serializability into the transactional system.

In theory, the anomalies of any isolation level, including snapshot isolation, could be dynamically detected at runtime by verifying the dependency graph [1, 2]. However, these approaches are very expensive for practical implementations. Cahill et. al [8] identify some low-granularity patterns that manifest in non-serializable executions of snapshot isolation. The verification of the patterns has, therefore, lower overhead compared to that of dependency graph. It, however, allows for false positives, which further lowers the concurrency level due to unnecessary aborts. Our lock-free implementation of write-snapshot isolation has the same overhead as lock-free implementation of snapshot isolation.

All the above approaches that try to serialize snapshot isolation inherit the unnecessary abort problem of write-write conflict avoidance. Write-snapshot isolation also unnecessarily aborts some transactions in read-write conflict avoidance. This is up to experimental results to show that which approach offers a higher level of concurrency. Our experiments in Section 6 show that the concurrency level offered by write-snapshot isolation under the mixed workload is comparable to that of snapshot isolation.

In an effort to generalize the ANSI isolation levels [3], Adya et. al. [1, 2] define anti-dependency for a transaction that writes a newer version of a value read by another transaction. The different isolation levels are, then, defined by avoiding certain cycles on the dependency graph. Although anti-dependency is not equivalent to read-write conflict defined in this paper, it was meant to capture the read-write conflicts in the dependency graph [1]. We define a read-write conflict only when they have rw-temporal overlap, while anti-dependency definition does not restrict the start and end of transactions. Moreover, anti-dependency makes sense when it is used to detect cycles on the dependency graph, where we have a perfect but expensive serializability without any false positive. In contrast, read-write conflict detection of this paper is a practical approach that is used without the global dependency graph of transactions and could, therefore, unnecessarily abort some transactions, similarly to the write-write conflict detection of snapshot isolation.

### 7.2 Implementations of Snapshot Isolation in Large-scale Data Stores

Percolator [24] takes a lock-based, distributed approach to implement snapshot isolation on top of BigTable. Percolator adds two extra columns to each column family: lock and write. Each transaction performs its writes directly into the main table. The write column is used to store commit timestamps. The lock columns simplify the write-write conflict detection since the two-phase commit algorithm run for each transaction avoids writing into a locked column. If a reading transaction finds the column locked, it has to check the status of the transaction that has locked the column. For this purpose, Percolator uses the state of a predefined modified entry by the transaction. The reading transaction, therefore, has to send a query to the server that maintains that particular entry.

Although using locks simplifies the write-write conflict detection, the locks held by a failed or slow transaction prevent the others from making progress until the full recovery from the failure. Moreover, maintaining the lock column as well as responding the queries about a transaction status coming from reading transactions puts extra load on data servers. To alleviate this extra load, Percolator [24] was forced to use heavy batching of messages sent to data servers, which inflicted a nontrivial, multi-second delay on transaction processing. We presented a lock-free implementation of write-snapshot isolation that does not suffer from the problems of using locks, and further provides serializability, the feature that snapshot isolation is missing.

Similar to Percolator, Zhang and Sterck [26] use the HBase data servers to store transactional data for snapshot isolation. However, the transactional data are stored on some separate tables. Even the timestamp oracle is a table that stores the latest timestamp. The benefit is that the system can run on bare-bone HBase. The disadvantage, however, is the low performance due to the many more accesses to the data servers to maintain the transactional data. Our approach provides serializability with a negligible overhead on data servers. ecStore [25] also provides snapshot isolation. To detect write-write conflicts, it runs a two-phase commit algorithm among all participant nodes, which has a scalability problem for general workloads.

### 7.3 Implementations of Serializability in Large-scale Data Stores

To achieve scalability, MegaStore [4], ElasTras [13], and G-Store [14] rely on partitioning the data store, and provide ACID semantics within partitions. The partitions could be created statically, such as in MegaStore and ElasTras, or dynamically, such as in G-Store. However, ElasTras and G-Store have no notion of consistency across partitions and MegaStore [4] provides only limited consistency guarantees across them. ElasTras [13] partitions the data among some transaction managers (OTM) and each OTM is responsible for providing consistency for its assigned partition. There is no notion of global serializability. In G-Store [14], the partitions are created dynamically by a *Key Group* algorithm, which essentially labels the individual rows on the database with the group identifier.

MegaStore [4] uses a write-ahead log to synchronize the writes within a partition. Each participant writes to the main database only after it successfully writes into the write-ahead log. Paxos is run between the participants to resolve the contention between multiple writes into the write-ahead log. Although transactions across multiple partitions are supported with an implementation of the two-phase commit algorithm, the applications are discouraged from using that due to performance issues.

Similar to Percolator, Deuteronomy [22] uses a lock-based approach to provide ACID. In contrast to Percolator where the locks are stored in the same data tables, Deuteronomy uses a centralized lock manager (TC). Furthermore, TC is the portal to the database and all the operations must go through it, making it the bottleneck for scalability. This leads to a low throughput offered by TC [22]. On the contrary, our approach is lock-free and can scale up to 50K TPS (500K write operations per second). Moreover, the data is accessed directly through HBase servers and in contrary to Deuteronomy do not go through the status oracle.

## 8. Concluding Remarks

In this paper, we contrasted read-write conflict with write-write conflict that is targeted by snapshot isolation. We proved that read-write conflict detection has the advantage of being serializable, the precious feature that snapshot isolation is missing. We showed that, similarly to snapshot isolation, write-snapshot isolation does not abort read-only transactions, which comprise the majority of transactional traffic. We then presented a new isolation level, write-snapshot isolation, which checks for read-write conflicts instead of write-write conflicts. Perhaps, the most important advantage of write-snapshot isolation is its simplicity for it efficiently adds serializability to a lock-free implementation of snapshot isolation by the slightest changes.

We showed that in a centralized, lock-free scheme of transactional support, which is suitable for large-scale data stores, the overhead of implementing both snapshot isolation and write-snapshot isolation is comparable. The offered level of concurrency highly depends on the particular workload that the application generates. The experimental results showed that snapshot isolation and write-snapshot isolation offer a comparable level of concurrency under a mixed, synthetic workload. The open source release of our implementation, Omid, is available to public [7], and can be tried out on future real-world transactional applications that will operate on top of distributed data stores.

## A. Implementation Details

Here, we briefly present our implementation of the status oracle that is covered in our previous works [17, 20]. The timestamps are obtained from a timestamp oracle integrated into the status oracle. The two main concerns related to the centralized scheme of status oracle are (i) *efficiency*, as the status oracle could potentially be a performance bottleneck,

---

[7] https://github.com/yahoo/omid

---

and (ii) *reliability*, as the status oracle could be a single point of failure.

Our implementation of the status oracle deployed on a simple dual core machine scales up to 50K TPS (where each transaction modifies 10 rows in average). To achieve this scale, the status oracle services requests from memory: it does not require a read from a hard disk to commit a transaction. However, to detect conflicts Line 2 of Algorithm 3 requires the commit timestamp of all the rows in the database, which does not fit in memory. To address this issue, the status oracle keeps only the state of the last NR committed rows that fit into the main memory, but it also maintains $T_{max}$, the maximum timestamp of all the removed entries from memory. Algorithm 3 shows the status oracle procedure to process commit requests.

---

**Algorithm 3** Commit request: {commit, abort}

1: for each row $r \in R$ do
2:     if *lastCommit*($r$) /= null then
3:         if *lastCommit*($r$) > $T_s(txn_i)$ then
4:             return abort;
5:         end if
6:     else
7:         if $T_{max}$ > $T_s(txn_i)$ then
8:             return abort;
9:         end if
10:    end if
11: end for
                                          ◁ Commit $txn_i$
12: for each row $r \in R$ do
13:    *committed*($r$)$T_s(txn_i) \rightarrow T_c(txn_i)$
14: end for
15: return commit

---

Line 8 pessimistically aborts the transaction, which means that some transaction could unnecessarily abort. It is not a problem if $T_{max} - T_s(txn_i) \gg MaxCommitTime$. Assuming 8 bytes for unique identifiers, we estimate the required space to keep a row data, including row identifier, start timestamp, and commit timestamp, at 32 bytes. Assuming 1 GB of memory, we can fit data of 32M rows in memory. If each transaction modifies 8 rows on average, then the rows for the last 4M transactions are in memory. Assuming a maximum workload of 80K TPS, the row data for the last 50 seconds are in memory, which is far more than the average commit time, *i.e.*, hundreds of milliseconds.

To check if a read version in the read snapshot of a transaction, we need access to the commit timestamp of the transaction that has written the version. The algorithm for performing this check is also changed to take into account the value of $T_{max}$. We refer the readers to our previous work [20] for more details. To reduce the load of performing this check on the status oracle, a read-only copy of the commit timestamps could be maintained in (i) data servers, beside the actual data [20], or (ii) the clients [17]. The results reported in this paper are produced using the latter approach.

To provide reliability for the in-memory data, the status oracle persists commit data into a write-ahead log. In this way, if the status oracle server fails, the same status oracle after recovery, or another fresh instance of the status oracle could still recreate the memory state from the write-ahead log and continue servicing the commit requests. The write-ahead log is also replicated across multiple remote storage devices to prevent loss of data after a storage failure. Writing into multiple remote machines could be very expensive and it is important to prevent it from becoming a bottleneck. We use Bookkeeper for this purpose, which could efficiently perform up to 20,000 writes of size 1028 bytes per second into a write-ahead log. Since status oracle requires frequent writes into the write-ahead log, multiple writes could be batched with no perceptible increase in processing time. With a batching factor of 10, BookKeeper is able to persist data of 200K TPS. The write of the batch to BookKeeper is triggered either by batch size, after 1 KB of data is accumulated, or by time, after 5 ms since the last trigger.

## Acknowledgments

We thank Russell Sears, the anonymous reviewers, and our shepherd, Maurice Herlihy, for the useful comments. This work has been partially supported by the Cumulo Nimbo project (ICT-257993), funded by the European Community.